\documentclass[a4paper,11pt]{article}
\usepackage[noblocks]{authblk}
\usepackage{here}
\usepackage{subfig}
\usepackage[pdftex]{graphicx}
\usepackage{color}
\usepackage{braket}
\usepackage{amssymb, amsmath}
\usepackage{feynmf}
\usepackage{bm}
\usepackage{url}
\usepackage{slashed}
\usepackage{caption}
\usepackage{jcappub}
\usepackage[compat=1.1.0]{tikz-feynhand}
\usepackage{hyperref}
\usepackage[samesize]{cancel}
\usepackage{fancyhdr}

\newcommand{\be}{\begin{eqnarray}}
\newcommand{\ee}{\end{eqnarray}}

\newcommand{\beq}{\begin{equation}}
\newcommand{\eeq}{\end{equation}}
\newcommand{\beqa}{\begin{eqnarray}}
\newcommand{\eeqa}{\end{eqnarray}}

\newcommand{\lmk}{\left(}
\newcommand{\rmk}{\right)}
\newcommand{\lkk}{\left[}
\newcommand{\rkk}{\right]}
\newcommand{\lnk}{\left\{}
\newcommand{\rnk}{\right\}}
\newcommand{\ri}{{\rm i}}

\newcommand{\la}{\langle}
\newcommand{\ra}{\rangle}

\begin{document}

\title{\boldmath Chiral gravitational effect in time-dependent backgrounds}
\author[a]{Kohei Kamada}
\author[a,b]{Jun'ya Kume}
\author[a]{and Yusuke Yamada}
\affiliation[a]{Research Center for the Early Universe (RESCEU), Graduate School of Science, The University of Tokyo, Hongo 7-3-1
Bunkyo-ku, Tokyo 113-0033, Japan}
\affiliation[b]{Department of Physics, Graduate School of Science,
The University of Tokyo, Hongo 7-3-1
Bunkyo-ku, Tokyo 113-0033, Japan}

% e-mail addresses: one for each author, in the same order as the authors
\emailAdd{kohei.kamada@resceu.s.u-tokyo.ac.jp}
\emailAdd{kjun0107@resceu.s.u-tokyo.ac.jp}
\emailAdd{yamada@resceu.s.u-tokyo.ac.jp}

\subheader{{\rm RESCEU-3/21}}

\abstract{Gravitational counterpart of the chiral magnetic effect, which is referred as the chiral gravitational effect, can also be of interest in a cosmological setup. In this study, we investigate this effect in the time-dependent chiral asymmetric fermion background and in the expanding spacetime by formulating
the effective action of gravitational waves. We also analyze the anomaly equation to see how the backreaction from %the
 gravitational waves to %the
 thermal chiral plasma occurs. 
We find that the non-trivial time dependence of chiral chemical potential, which can be induced
in some scenarios of baryogenesis, is the key ingredient of the chiral gravitational effect. It turns out that the ``memory" of the effect is imprinted on the high frequency gravitational waves propagating in the plasma. 
Cosmological implications and potential effects on the gravitational wave observation are briefly discussed.}
\maketitle
\flushbottom

\section{Introduction}
Parity violating effects related to quantum anomalies have been discussed in various fields of physics, from condensed matter to cosmology. One of the representative examples is the ``Chiral Magnetic Effect"(CME)~\cite{Vilenkin:1980fu, Nielsen:1983rb, Alekseev:1998ds, Son:2004tq, Fukushima:2008xe}, which arises in the theory with a U(1) gauge symmetry. The CME is a phenomenon where the magnetic field induces the electric current parallel to the field itself, under the presence of the asymmetry between left- and right-handed fermions %characterized by the chiral chemical potential $\mu_5$
or the chiral asymmetry, which give rise to the distinct transport phenomena. 
Since the chiral anomaly is ubiquitous in quantum field theory, 
it is expected to appear in a broad range of energy scales, such as in the Weyl semimetals~\cite{Zyuzin:2012tv,Goswami:2012db,Chen:2013mea,Basar:2013iaa,Hosur:2013kxa,Landsteiner:2013sja,Chernodub:2013kya}, 
relativistic heavy ion collisions~\cite{Kharzeev:2010gd,Burnier:2011bf,Hongo:2013cqa,Yee:2013cya,Hirono:2014oda,Huang:2015oca,Kharzeev:2015znc,Shi:2017cpu}, or compact stars~\cite{Charbonneau:2009ax,Ohnishi:2014uea,Grabowska:2014efa,Kaminski:2014jda,Sigl:2015xva,Yamamoto:2015gzz,Masada:2018swb}. 
Moreover, in the hot early Universe, where strong magnetic fields~\cite{Grasso:2000wj} as well as chiral asymmetry~\cite{Campbell:1992jd} can exist, 
interesting phenomena can take place.
Since the current induced by the CME leads to the instability in the magnetic fields in chiral plasma,
known as the chiral plasma instability (CPI)~\cite{Joyce:1997uy,Akamatsu:2013pjd}, %the CME provides rich phenomenology to the primordial thermal plasma of the universe. For example,
it is, {\it e.g.}, ascribed as the possible origin of the primordial magnetic fields~\cite{Joyce:1997uy,Tashiro:2012mf}, 
which leads to the baryogenesis~\cite{Kamada:2018tcs} from the hypermagnetic helicity decay~\cite{Giovannini:1997gp,Giovannini:1997eg,Bamba:2006km,Fujita:2016igl,Kamada:2016eeb,Kamada:2016cnb}. 
Numerical magnetohydrodynamic simulations~\cite{Masada:2018swb,Schober:2017cdw} as well as lattice simulations~\cite{Figueroa:2019jsi,Mace:2019cqo} have recently been performed
to show how the CPI proceeds in thermal environments.
%On the other hand, this chiral plasma instability may cancel the helicity of the magnetic fields and the baryons produced in the early universe~\cite{Domcke:2019mnd}. 
Since the amplified electromagnetic fields contributes to
the transverse and traceless part of the energy momentum tensor, 
gravitational waves are emitted~\cite{Anand:2018mgf, Brandenburg:2021aln}, which is one of the interesting consequences 
of the effect. 
The primordial chiral asymmetry, the key ingredient in these processes, can be generated in well-motivated cosmological 
phenomena such as the grand unified theory (GUT) baryogenesis and axion inflation~\cite{Kamada:2018tcs,Domcke:2018eki,Domcke:2019mnd,Domcke:2020quw}.

While the cosmological application of the CME is now being investigated in detail, 
its gravitational counterpart referred as ``chiral gravitational effect''~\cite{Sadofyev:2017zqc} has not been extensively investigated so far. In the case of chiral gravitational effect, 
the graviton (the transverse and traceless part of the metric perturbation) receives parity-violating vacuum polarization from the 
chiral asymmetry in the system, which leads to the anomalous contribution to the energy momentum tensor.\footnote{See also Ref.~\cite{Avdoshkin:2020han} for the study on the chiral gravitational effect from the holographic point of view.} 
Note that parity-violating effect is ascribed to the ``gravitational (chiral) anomaly''~\cite{AlvarezGaume:1983ig}\footnote{The chiral vortical effect partially originates from the gravitational chiral anomaly on the vector part of the metric perturbation~\cite{Landsteiner:2011cp}.} in this case. 
Applications and discussions of this anomaly in cosmology have mainly focused on the study of leptogenesis, in which it connects the lepton chiral charge to the helicity of gravitational waves. This is realized in pseudo-scalar inflation models with the gravitational Chern-Simons term, where the chiral gravitational waves and the primordial lepton asymmetry are simultaneously generated during inflation through this anomaly~\cite{Alexander:2004us, Lyth:2005jf, Fischler:2007tj, Kamada:2019ewe}.\footnote{See also Ref.~\cite{Kawai:2017kqt} for the case where the Gauss-Bonnet term contribution is incorporated in this model.}
In this simple setup, unfortunately, the resultant lepton asymmetry is too small to explain matter anti-matter asymmetry of the Universe~\cite{Fischler:2007tj, Kamada:2020jaf}, though.\footnote{Gravitaional leptogenesis can be successful in the models where the inflaton couples to the Chern-Pontryagin density of the gauge fields~\cite{Noorbala:2012fh, Maleknejad:2014wsa, Caldwell:2017chz, Papageorgiou:2017yup, Adshead:2018doq}. See also Ref.~\cite{Adshead:2017znw} where the effect of lepton number violation due to the Majorana neutrino mass has been discussed.}
On the other hand, the chiral gravitational effect should also be able to cause interesting phenomena in the primordial thermal plasma with net chiral asymmetry, in a similar way to the CPI.  
Namely, the anomalous energy momentum tensor induced by the chiral gravitational effect~\cite{Manes:2012hf,Jensen:2012kj} modifies the propagation of gravitational waves, and it 
in turn affects the fermion background through the anomaly. %\KK{KK:[Strictly speaking, the backreaction itself is not the CPI.]}
Thus, one can expect non-trivial interplay between the chiral imbalance of the plasma and the gravitational wave in the early Universe.  

Before investigating the chiral gravitational effect in realistic cosmological setups, one needs to extend the formulation in the previous study. One of the pioneer works on the chiral gravitational effect~\cite{Sadofyev:2017zqc} considered this effect in the static thermal fermion background with a chiral asymmetry, based on Refs.~\cite{Manes:2012hf,Jensen:2012kj} where an anomalous dispersion relation of the gravitational waves in such a background is derived. 
To investigate this effect in the primordial chiral thermal plasma, one needs to go beyond the static background
since the chiral asymmetry characterized by the chemical potential $\mu_5$ exhibits a non-trivial time dependence at 
the generation and decay. 
The former typically takes place at some baryogenesis scenarios such as the GUT baryogenesis~\cite{Kamada:2018tcs,Domcke:2020quw}, 
whereas the latter happens when the electron Yukawa interaction gets in equilibrium~\cite{Campbell:1992jd}.
Note that unlike vector fields, gravitons are %is
not conformally invariant, and hence the dilution due to the cosmic expansion
also affects the phenomena induced by the chiral gravitational effect.
In this study, we provide the effective field theoretic view point to incorporate the time dependence of the fermion background with a chiral imbalance and that of the background geometry into the chiral gravitational effect.

A theoretical aspect of the CME is that the effect of the chiral anomaly on the gauge field dynamics can be understood as the induced Chern-Simons term in the effective Lagrangian, which can be obtained by integrating out the fermions coupled to
finite chiral chemical potential~\cite{Redlich:1984md}. In this effective Lagrangian, an asymmetric state of thermal plasma can be expressed by introducing a spurious scalar field $\theta(x)$ which anomalously couples to the Chern-Pontryagin density~\cite{Kharzeev:2009mf,Boyarsky:2015faa}. This formulation of the effective field theory  studied in the CME can enable us to incorporate the time-dependent behavior of the background and 
we here apply the similar procedure to the chiral gravitational effect. 
Note that a similar discussion has been studied in Ref.~\cite{Barrie:2017mmr} where the authors consider the effect of an asymmetric fermion background, namely the cosmic neutrino background, to the propagating gravitational wave during the matter-dominated era, through the gravitational Chern-Simons term in the effective Lagrangian.
Nevertheless, we should note that there are non-local thermal and chemical potential corrections found in Refs.~\cite{Manes:2012hf,Jensen:2012kj}, which cannot be described within the local Lagrangian approach. We will focus on relatively high momentum gravitational waves, for which non-local effects are negligible or subdominant.

To investigate comprehensively the evolution of the chiral plasma and gravitational waves under the presence of the gravitational anomaly, we need to take into account the backreaction on the chiral plasma, which is another subject of this study.  
As a first step, we analyze simple toy models in the following way:
First, we derive the gravitational wave solutions perturbatively corrected by the gravitational Chern-Simons term both in Minkowski spacetime and Friedmann-Robertson-Walker (FRW) background, which clarify how the chiral gravitational effect affects the propagation of gravitational waves.
We here assume that background gravitational waves have already existed in the system, 
which is motivated by the stochastic gravitational wave background from such as inflation~\cite{Grishchuk:1974ny,Starobinsky:1979ty}, 
preheating~\cite{Khlebnikov:1997di, Easther:2006gt, Easther:2006vd, Dufaux:2007pt}, and first-order phase transitions~\cite{Kosowsky:1991ua, Kosowsky:1992rz, Nicolis:2003tg, Grojean:2006bp}.
Then we consider its backreaction to the chiral plasma through the anomaly equation for models in both Minkowski and FRW backgrounds. %the both spacetime.
Note that we restrict ourselves into the region where the chiral chemical potential and the effect of the Chern-Simons term is sufficiently small, to avoid the ghost instability, which will be explained later.

One of the main results of our investigation, explicitly shown in the Minkowski spacetime analysis, is that the propagation of gravitational wave is affected only when the chiral chemical potential $\mu_5$ has non-trivial time dependence. 
Such a behavior of the chiral gravitational effect is in contrast with the case of CME where the instability is triggered simply by the presence of non-vanishing constant $\mu_5$.\footnote{When one considers a time-dependent pumping of electrons in asymmetric Weyl semimetals, the anomalous transport is characterized by the time derivative of $\mu_5$ even in the case of CME~\cite{Kharzeev:2016mvi}.} We emphasize that the expansion of the Universe non-trivially contributes to the chiral gravitational effect since gravitons are not conformal. As we will see, the expansion leads to effectively time dependent chemical potential, which triggers non-trivial interplay between the gravitational waves and the chiral chemical potential through the chiral gravitational effect. We also demonstrate that the time dependence of chemical potential  generically leaves the imprints of the primordial chirality violation on gravitational waves. Then we show that the backreaction from gravitational waves to chiral plasma is typically small compared to the dominant part in chemical potential dynamics. %But 
However, when the dominant part eventually decays, the ``imprints'' at an earlier time can become important for late time behavior of chemical potential. There, the backreaction leads to the oscillatory conversion between the chiral chemical potential and the helicity of gravitational waves driven by the external chirality-changing process. 

The rest of the paper is organized as follows. In Sec.~\ref{CS}, we make a brief review of the derivation of the gravitational Chern-Simons term in the effective Lagrangian arising from chiral asymmetric fermion background. Then we discuss its general physical property that affects the propagation of the gravitational waves as a consequence of chiral gravitational effect. By incorporating the induced Chern-Simons term into the equation of motion, we explicitly show  the corrected %calculate the propagation of 
gravitational waves in time-dependent chiral plasma for the Minkowski spacetime and evaluate its backreaction through the anomaly equation in Sec.~\ref{GW_sol}. 
In Sec.~\ref{br}, we consider the FRW spacetime to investigate the effect of cosmological expansion on the chiral gravitational effect. In the same way as in the Minkowski spacetime case, we solve the modified equation of motion and then evaluate the backreaction from gravitational waves to the chiral %ity of the 
 plasma through the gravitational anomaly. The final section is devoted to the summary and discussion.
Throughout this paper, we will use the convention in Ref.~\cite{Alexander:2009tp}, {\it e.g.}, $\eta_{\mu\nu}={\rm diag}(-1,1,1,1)$ and the Levi-Civita symbol $\epsilon^{\alpha\beta\gamma\delta}$ satisfying $\epsilon^{0123}=+1$, whereas the Levi-Civita tensor is given by $\hat{\epsilon}^{\alpha\beta\gamma\delta}=\frac{1}{\sqrt{-g}}\epsilon^{\alpha\beta\gamma\delta}$.

\section{Chiral gravitational effect in a dynamical background}\label{CS}
We first discuss the gravitational parity-violating effect in a chiral asymmetric fermion background, which is referred as the chiral gravitational effect in Ref.~\cite{Sadofyev:2017zqc}, focusing on the Minkowski spacetime background for simplicity. 
In order to consider this phenomenon in a cosmological setup, {\it e.g.} primordial thermal plasma of the Universe, one needs to incorporate the dynamical behavior of the chiral fermion background, 
associated with its generation 
mechanism and the chirality-flipping interactions. 
We will see that the dilution due to the cosmic expansion also affects 
the dynamics of gravitons when we consider the generalization to the FRW background.
 Such a dynamics has been taken into account in Ref.~\cite{Barrie:2017mmr} where  the effect of the cosmic neutrino background on the propagating gravitational waves through the induced gravitational Chern-Simons term is considered. We here give a brief review of the formulation in Ref.~\cite{Barrie:2017mmr} (see also Ref.~\cite{Assuncao:2018jkq}), which is consistent with Ref.~\cite{Manes:2012hf}, and re-examine it in the context of chiral gravitational effect 
for %having in mind 
the application to cosmology.

\begin{figure}[htbp]
  \centering
  \includegraphics[width=8cm,clip]{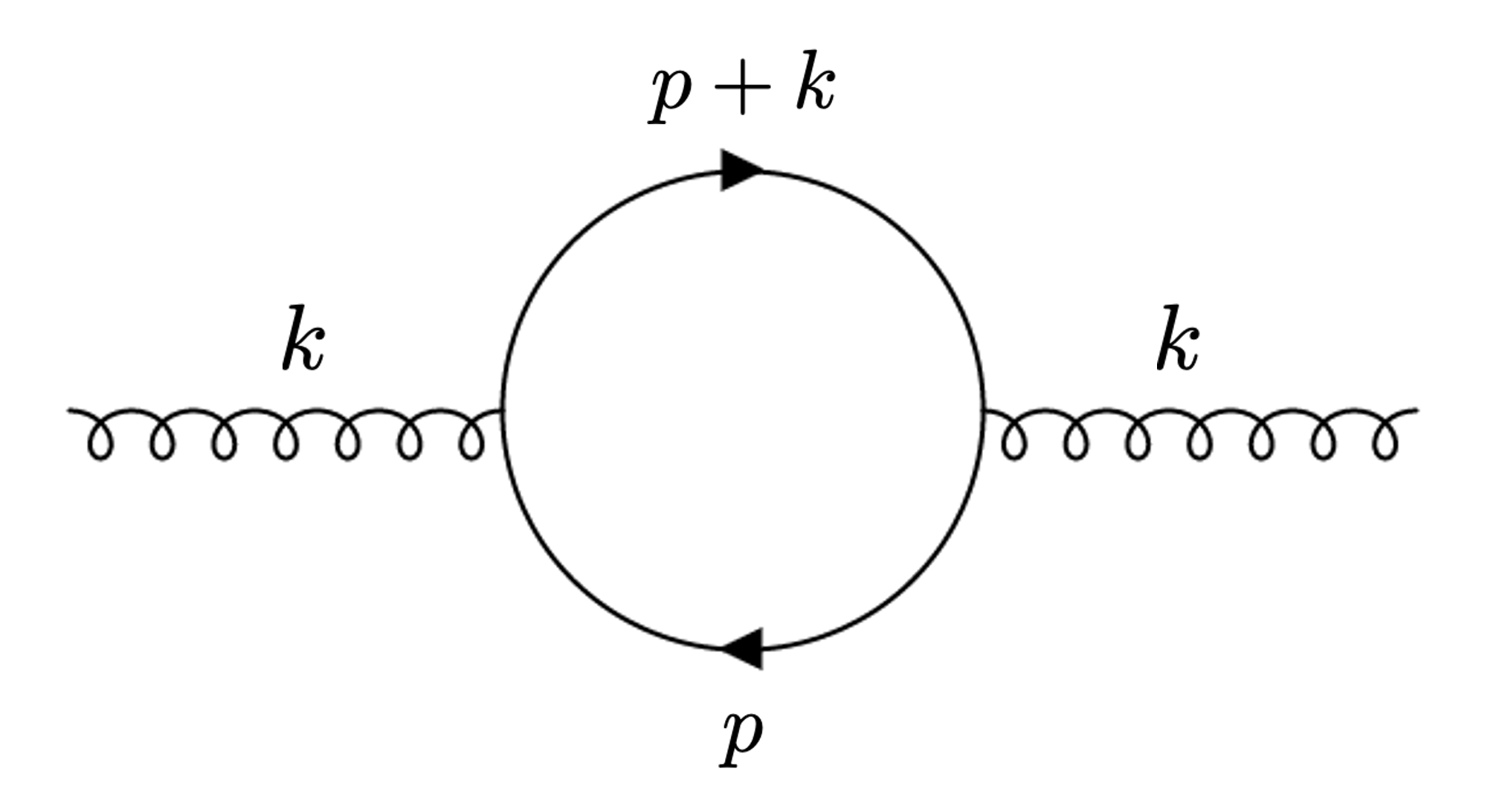}
    \caption{1-loop correction to the graviton propagator relevant to the induction of the gravitational Chern-Simons term.}
    \label{fig1}
\end{figure}

Let us consider the tensor perturbation (graviton) around the Minkowski spacetime $g_{\mu\nu} \simeq \eta_{\mu\nu} + h_{\mu\nu}$ with a chiral asymmetric fermion background.
The chiral asymmetry of the fermion system is characterised by the chiral chemical potential $\mu_5$ and is described by a quadratic term of fermions in the effective Lagrangian as 
\beq
S_{\mu}^\mathrm{eff} = \int d^4 x \mu_5\bar{\psi}\gamma^0\gamma^5\psi.\label{chemical pot}
\eeq
Throughout this paper, we focus on the regime where the typical momentum of the graviton, $k$, is larger than the background temperature $T$ and the chemical potential $\mu_5$, namely, $k \gg |\mu_5|, T$,\footnote{We will clarify the reason why we need to assume such a parameter region before the end of this section.} where the contribution from the induced Chern-Simons term dominates 
other contributions, {\it e.g.} %a
non-local thermal corrections, found in Ref.~\cite{Sadofyev:2017zqc,Manes:2012hf} in the anomalous responses. The term induced by the chiral asymmetry (Eq.~\eqref{chemical pot}) modifies the propagator of the fermion, which in turn gives the parity-violating contribution to the graviton polarization tensor $\Pi_{\mu\nu\rho\sigma}$ through the loop correction as depicted in Fig.~\ref{fig1}.
In the massless limit of the fermion $m \ll k$, the leading parity-violating correction to the polarization tensor from the fermion loop takes the following simple form in the momentum space~\cite{Barrie:2017mmr}:
\beq
\Pi_{\mu\nu\rho\sigma}^\mathrm{PV} = \mathrm{i} \frac{\mu_5}{192\pi^2}\lnk\epsilon_{\mu\rho\alpha0}k^{\alpha}[k_{\nu}k_{\sigma} - k^2\eta_{\nu\sigma}]+(\mu\leftrightarrow\nu)+(\rho\leftrightarrow\sigma)+(\mu\leftrightarrow\nu,\rho\leftrightarrow\sigma)\rnk,
\eeq
where $k$ denotes the momentum of the graviton. Here the higher order terms in $\mu_5$ and $k$ are neglected. % since we here are interested in the linear terms in $\mu_5$ which results in the gravitational Chern-Simons term. 
This loop correction can be rewritten in terms of the gravitational Chern-Simons term in 
the effective action up to the second order in $h_{\mu\nu}$ as %\KK{KK: [The sign seems to be opposite to Ref.~\cite{Assuncao:2018jkq} taking into account the definition of $b_\mu$ but let us rely on Ref.~\cite{Barrie:2017mmr}. In our study, the choice of left and right handed helicity is not important. But when we would like to compare realistic models of baryogenesis and observations of the helicity of the gravitational waves, we need to be careful on that.]} 
%Then we can rewrite this loop correction in terms of the effective action of the graviton as
\beq
\begin{split}
S_2^\mathrm{eff,PV}  &= -\frac{\mu_5}{192\pi^2}\int d^4x \epsilon_{\mu\rho\alpha 0}h^{\mu\nu}\partial^{\alpha}(\Box h^{\rho\sigma}\eta_{\nu\sigma} - \partial_{\nu}\partial_{\sigma}h^{\rho\sigma})\\
  &=\frac{\mu_5}{96\pi^2}\int d^4x K^{0},
\end{split}
\label{S2eff}
\eeq
where $K^0$ is the linearized 0-th component of the four dimensional Chern-Simons topological current,  $K^\mu \equiv \epsilon^{\mu\nu\rho\sigma} \Gamma^\alpha_{\nu\beta} ( \partial_\rho \Gamma^\beta_{\sigma \alpha}+ (2/3) \Gamma^\beta_{\rho\gamma} \Gamma^\gamma_{\sigma \alpha})$.

It is useful to introduce a spurious scalar field $\theta$ which is related to the chiral chemical potential as
\beq
\partial_{\mu}\theta = \lmk\frac{\mu_5}{192\pi^2}, 0, 0, 0\rmk.\label{scalar}
\eeq
By virtue of this field, the influence of the general background with the chiral asymmetry results in the form of gravitational Chern-Simons term as
\beq
S_\mathrm{eff}^\mathrm{PV} =  \int d^4x(\partial_{\mu}\theta)2K^{\mu} = -\int d^4x \sqrt{-g}\theta R\tilde{R},\label{gCS_f}
\eeq 
with the Chern-Pontryagin density being defined as
\beq
R\tilde{R} \equiv R_{\mu\nu\alpha\beta}\tilde{R}^{\mu\nu\alpha\beta} = -\frac{1}{2}\frac{\epsilon^{\alpha\beta\gamma\delta}}{\sqrt{-g}}R_{\alpha\beta\rho\sigma}R_{\gamma\delta\mu\nu}g^{\mu\rho}g^{\nu\sigma}.
\eeq
Note that here the induced parity-violating term is extended into the covariant expression. %so that it is straightforward to generalize to the spacetime-dependent chiral chemical potential and to the curved space time background such as the FRW spacetime.
Although the derivation of the effective action is based on the constant chemical potential background in the Minkowski spacetime, the covariant expression enables us to apply the effective action to more general situations where the chemical potential effectively has nontrivial time dependence and to the curved spacetime case such as the FRW spacetime. 
Such an extension would be reasonable since our approach can be understood as the effective field theory where the (spacetime) symmetry of the theory is spontaneously broken by the spurious field $\theta$. See also the related studies of the CME~\cite{Kharzeev:2009mf,Boyarsky:2015faa}, which also introduced the spurious axion field $\theta(t,{\vec x})$ coupled to the $F\tilde{F}$ term in order to represent the asymmetric state of the thermal plasma.\footnote{The ``spurious" scalar field $\theta$ actually appears as a collective excitation in the anomalous hydrodynamics. See Ref.~\cite{Monteiro:2014wsa} for the classical case and Ref.~\cite{Mottola:2019nui} for the quantum case.} 

Let us clarify the physical property of the induced Chern-Simons term by expanding the effective action around the Minkowski spacetime background. After taking the transverse and traceless gauge, the action can be expanded up to the second order in the metric perturbation $h_{ij}$ as
\beq
\begin{split}
  S_{\text{E-H}}+S_{\rm eff}^{\rm PV} \simeq \frac{M_{\rm Pl}^2}{8}\int d^4x & \left[\lnk\dot{h}^i_{\ j}\dot{h}^j_{\ i} - (\partial_kh^i_{\ j})(\partial^kh^j_{\ i})\rnk \right. \\ 
    & \left.+ \lmk\frac{8{\dot \theta}}{M^2_{\rm Pl}}\rmk\epsilon^{0ijk}\lnk\dot{h}^q_{\ i}\partial_j\dot{h}_{kq} - (\partial^rh^q_{\ i})\partial_j\partial_rh_{kq}\rnk \right], \label{S2}
\end{split}
\eeq
where $S_{\text{E-H}}$ denotes the Einstein-Hilbert term and the dot denotes the time derivative. 
This action leads to the equation of motion
\beq
\begin{split}
\ddot{h}_{\ i}^{j} - \nabla^2h_{\ i}^{j} &= -\frac{8}{M_{\rm Pl}^2}\epsilon^{jmn}\partial_m\lmk\dot{\theta}\ddot{h}_{in} + \ddot{\theta} \dot{h}_{in} - \dot{\theta}\nabla^2h_{in}\rmk\\
&=-\frac{\epsilon^{jmn}}{24\pi^2M^2_{\rm Pl}}\partial_m\lmk\mu_5\ddot{h}_{in} + \dot{\mu}_5 \dot{h}_{in} - \mu_5\nabla^2h_{in}\rmk,\label{eom_f}
\end{split}
\eeq
where we have used Eq.~\eqref{scalar} in the second line. We will study its consequences in the following sections.

Before studying the solutions of Eq.~\eqref{eom_f}, let us investigate the physical implications 
of the induced term.
The modification of the equation of motion in Eq.~\eqref{eom_f} reflects the fact that the Chern-Simons term in the effective action leads to the transverse-traceless part of the induced parity-violating energy momentum tensor in the asymmetric fermion background as
\begin{eqnarray}
T^{(\mathrm{ind})}\ ^j_{\ i}
&\simeq& 4\epsilon^{jmn}\lnk\dot\theta\partial_m\lmk\partial_t^2 -\nabla^2\rmk+ \ddot{\theta}\partial_m\partial_t\rnk h_{in}.\label{CGE}
\end{eqnarray}
From this expression, one can understand that the effect is analogous to the CME where the electric current is induced by the external magnetic field. In the present case, the external gravitational field applied to the chiral asymmetric system  induces the energy momentum tensor, which is the ``current" for the diffeomorphism. Therefore, it is reasonable to call this phenomenon as the chiral gravitational effect. 

The chiral gravitational effect in a static thermal background has been studied in Ref.~\cite{Sadofyev:2017zqc}, where the authors particularly focused on the anomalous corrections to the dispersion relation and studied the properties of the corrections in various limits of parameters. The analysis is based on  the gravitational response function derived in literature~\cite{Manes:2012hf,Jensen:2012kj}. The term proportional to $\mu_5$ in Eq.~\eqref{CGE} actually coincides with the one derived in the previous study~\cite{Manes:2012hf} in the static limit of the gravitational field. On the other hand, the latter term proportional to $\dot{\mu}_5$ appears only when we 
introduce the effective action with the induced Chern-Simons term (Eq.~\eqref{gCS_f})
that allows the time-dependent %background
chiral chemical potential.
%express the \KK{background} chemical potential with nontrivial time dependence by introducing the Chern-Simons term in the effective action as Eq.~\eqref{gCS_f}. 
In this way, the dynamical behavior of the chiral fermion background has been incorporated into the chiral gravitational effect. Although we here demonstrate it in
the Minkowski spacetime case, a curved background geometry is also expected to give non-trivial contributions to the chiral gravitational effect due to the non-conformal nature of the graviton as we will see in Sec.~\ref{br}. 

We would also like to mention the difficulty in the Lagrangian description of the thermal correction as well as the higher order corrections from the non-zero chemical potential. The thermal and chemical potential corrections to the two point function found in Refs.~\cite{Manes:2012hf,Jensen:2012kj} contain non-local contributions as well as local one. The latter corresponding to the gravitational Chern-Simons term can be written in a covariant form as we have seen. 
On the other hand, it is in general difficult to rewrite the former in such a form. Indeed, 
in Ref.~\cite{Assuncao:2018jkq}, %the authors show 
the local as well as non-local Lagrangian terms including thermal corrections are shown
at the leading order of $\mu_5$ in a non-covariant form, 
but a covariant Lagrangian description that takes into account all the contributions 
found in Refs.~\cite{Manes:2012hf,Jensen:2012kj}  has not been obtained, if ever. The covariant (local) Lagrangian is an important guiding principle for us to extend the effective Lagrangian~\eqref{gCS_f} to that in more general backgrounds. 
It is desirable if we would have a covariant form of the non-local contributions to 
apply such a prescription, but at present we have not reached at it yet.
This is the reason why we discuss the parameter region $k \gg  T, |\mu_5|$ so that the local contribution dominates over non-local ones. We should also emphasize that even in a different parameter region, where non-local terms dominate the graviton dynamics, the local contribution is present and the effects we discuss in this paper would remain.% even though they become subdominant.

Before closing this section, we would like to comment on the possible issues associated with the gravitational anomaly term (Eq.~\eqref{S2}). As we can see from Eq.~\eqref{eom_f}, the additional term contributes to the kinetic term  (or the dispersion relation) if we Fourier-transform the expression. In particular, such corrections depend on the momentum. Since this contribution to two point function is positive or negative depending on the chirality, either left or right modes with large enough momentum $k$ may become ghost-like.\footnote{This issue is one of the obstacles in gravitational leptogenesis with the gravitational anomaly term, see {\it e.g.} Ref.~\cite{Kamada:2019ewe}.} Therefore, one should introduce a cut-off scale to the theory above which the effective theory is broken down. In our following analysis, we will treat the correction perturbatively, or equivalently, we consider the energy and momentum range to be much below the cut-off scale, which is reasonable to avoid the appearance of the ghost-like mode mentioned above.\footnote{Our treatment of the correction is different from that in Ref.~\cite{Barrie:2017mmr}, where the authors discuss the constraint {\it e.g.} on the photon production caused by the ghost-like graviton. In order to avoid the subtlety associated with the treatment of such unstable modes, we simply consider the momentum range where  the kinetic terms of both the left and right graviton modes are positive.}

\section{Gravitational birefringence and the backreaction to chiral plasma in the Minkowski background}\label{GW_sol}

Now we are ready to investigate 
%In order to see 
how the induced energy momentum tensor contributes to the propagation of gravitational waves. For this purpose, in Sec.~\ref{GW_f}, we shall solve the equation of motion for %gravitational waves
gravitons by taking the gravitational Chern-Simons term as perturbation in the Minkowski spacetime. We will see that $\dot{\mu}_5$ dependence in Eq.~\eqref{CGE}, which is absent in the case of the CME, causes the gravitational birefringence. With the corrected gravitational wave solutions, we also analyze the anomaly equation in order to take into account the backreaction of gravitational waves to chiral plasma in~Sec.~\ref{br_f}. Such a comprehensive study of this system reveals the nontrivial interplay between gravitational waves and chiral plasma as we will present.

%We should note that the response related to the background temperature derived in Ref.~\cite{Manes:2012hf} has not been included in our formulation though it becomes dominant in the low momentum regime $k \lesssim |\mu_5|, T$. We will come back to this point in the final section.

\subsection{Propagation of gravitational waves}\label{GW_f}
We start with a toy model in the Minkowski spacetime background. % to illustrate the interplay between chemical potential and gravitational wave through the gravitational anomaly.
%For the concreteness, 
Specifically, we consider gravitational waves propagating in the $z$-direction throughout this section. Let us define the circular polarization tensors for the right and left-handed modes
as
\beq
p_{ij}^{A} \equiv \frac{1}{\sqrt{2}}(p_{ij}^{+} + \lambda_A p_{ij}^{\times}) = \frac{1}{\sqrt{2}}
\begin{pmatrix}
1 & \ri \lambda_A &0\\
\ri \lambda_A& -1 &0\\
0 &0 &0\\
\end{pmatrix}, \quad \text{with} \quad \lambda_{\rm R} = +1, \lambda_{\rm L} = -1, 
\eeq
which satisfy the following relations:
\begin{equation}
\begin{split}
  p_{ij}^\mathrm{R}p^{ij\mathrm{R}} &= p^\mathrm{L}_{ij}p^{ij\mathrm{L}} = 0,\\
  p_{ij}^\mathrm{R}p^{ij\mathrm{L}} &= 2,\\
  \epsilon^{m3j} p^A_{ij} &= -\ri\lambda_Ap^{m \  A}_{\ \ i}. %\ (\text{with} \ ).
\end{split} \label{cptr}
\end{equation}
Expanding the metric perturbation as $h_{ij} = \sum_{A = R,L}h_A(t,z)p_{ij}^A$, we can rewrite the equation of motion~\eqref{eom_f} as
\beq \label{eom_ff}
(\partial_t^2 - \partial_z^2)h_A(t,z) = \frac{{\rm i}\lambda_A}{24\pi^2M_{\rm Pl}^2}\partial_z(\mu_5\ddot{h}_A(t,z)+\dot\mu_5\dot{h}_A(t,z)-\mu_5\partial_z^2h_A(t,z)),
\eeq
so that the right- and left-handed modes are decoupled from each other.
Here $h_L(t,z)$ and $h_R(t,z)$ are related to each other as $h_{\rm R}(t,z) = h_{\rm L}^*(t,z)$, which follows from the reality condition. 

Let us investigate the correction to the plane wave solution of the form $h_A(t,z) = h_A(t)e^{{\rm i}kz} + h^*_{\bar A}(t) e^{-{\rm i}kz} $, 
where ${\bar A}$ is defined as $\bar{\mathrm{R}} = \mathrm{L}$ and $\bar{\mathrm{L}}=\mathrm{R}$. 
%generated by a dynamical mechanism,
From Eq.~\eqref{eom_ff}, %the equation of motion for $h_A(t)$ reads
$h_A(t)$ should satisfy
\beq
\begin{split}
  \lmk\frac{d^2}{dt^2} + k^2\rmk h_A(t) &= -\frac{\lambda_A k}{24\pi^2M_{\rm Pl}^2}\lnk\mu_5\lmk\frac{d^2}{dt^2} + k^2\rmk+\dot\mu_5\frac{d}{dt}\rnk h_A(t)\\
  &\equiv -\frac{\lambda_A k}{24\pi^2M_{\rm Pl}^2}\hat{F}_k(t)h_A(t).\label{eom_f2}
\end{split}
\eeq
One can see that the first two terms in the right-hand side of Eq.~\eqref{eom_f2}, 
coming from the induced Chern-Simons term, 
is proportional to its left-hand side originally contained in the Einstein-Hilbert term. 
This suggests that the Chern-Simons term may lead to the strong coupling and the ghost instability for the graviton when the effective coupling constant, $k |\mu_5|/24 \pi^2 M_\mathrm{Pl}^2$, 
is too large, 
%becomes compatible with that from the Einstein-Hilbert term, 
as we mentioned in the previous section. Such an instability is inevitable for large momentum and we cannot predict the dynamics of the system at 
the scale higher than the emergent strong coupling scale, 
$k> 24 \pi^2 M_\mathrm{Pl}^2/|\mu_5|$.
%Practically 
This instability is harmless if %unless we cut-off the energy range since the coefficient of $\partial_t^2h_A$ depends on the momentum and becomes negative for one of the polarization modes with large $k$. 
%Thus, we put a conservative constraint on the amplitude of the chemical potential as 
\beq
|\mu_5| \ll 24\pi^2 M_{\rm Pl}\label{constraint}
\eeq
is satisfied since the positivity of the coefficient of $d^2/dt^2$ in Eq.~\eqref{eom_f2} is ensured and the emergent scale lies in the strong coupling scale 
of the Einstein-Hilbert action, that is, the Planck scale~$M_{\rm Pl}$. 
We shall take Eq.~\eqref{constraint} as a conservative constraint on the chiral chemical potential.
%This ensures the positivity of the coefficient of $d^2/dt^2$ in eq.~\eqref{eom_f2} for the modes satisfying $k \leq M_{\rm Pl}$. %In the following, we also assume that $|\dot{\mu}_5|$ is much smaller than $M_{\rm Pl}^2$ for simplicity.

Under such an assumption on the parameter space, we can safely treat the contribution of the Chern-Simons term in the right-hand side of Eq.~\eqref{eom_f2} as a perturbation. 
We incorporate the effect of the gravitational Chern-Simons term perturbatively by using the Green function method. Here, the Green function is defined as the solution to 
\beq
 \lmk\partial_t^2 + k^2\rmk G(t,t') = \delta(t - t').
\eeq
The zero-th order solution $h_A^{(0)}(t)$ is nothing but the solution to Eq.~\eqref{eom_f2} for $\mu_5 = \dot{\mu_5} = 0$. Then, the first order correction to the zero-th order solution can be %calculated 
evaluated as
\beq \label{sol0}
h^{(1)}_A(t) = h^{(0)}_A(t)-\lambda_A\frac{k}{24\pi^2M_{\rm Pl}^2}\int_{-\infty}^{\infty}dt'G(t,t')\hat{F}_k(t')(h^{(0)}_A(t')),
\eeq
where $G(t,t') = \Theta(t-t')\sin k(t-t')/k$ is the explicit form of the Green function in this case with $\Theta(t)$ denoting the Heaviside step function. If we take the zero-th order plane wave solution to be that for the left-handed polarization mode, $h^{(0)}_\mathrm{L}(t) = A e^{-\mathrm{i}kt}$, Eq.~\eqref{sol0} reads 
\beq
h_{\rm L}^{(1)}(t) = A\lmk1+\frac{k}{48\pi^2M_{\rm Pl}^2}\Delta\mu_5(t)\rmk e^{-\ri kt}-\frac{Ak}{48\pi^2M_{\rm Pl}^2}\lmk\int_{-\infty}^tdt'\dot{\mu_5}(t')e^{-2\ri kt'}\rmk e^{\ri kt},\label{sol}
\eeq
where $\Delta\mu_5(t)=\mu_5(t)-\mu_5(-\infty)$.
As one can see from Eq.~\eqref{sol}, gravitational waves feel the {\it time variation} of the chiral chemical potential during the propagation but not the non-vanishing chiral chemical potential itself. If $\dot{\mu}_5 > 0$ for a while, as an example, the positive frequency mode is amplified and the negative frequency mode is simultaneously generated for the left-handed polarization. 
Note that the anomaly corrected solution for the right-handed polarization mode with a positive frequency, $h_\mathrm{R}(t) = A e^{-\ri kt}$, is obtained by replacing $\mu_5$ to $-\mu_5$ in Eq.~\eqref{sol} and results in the opposite correction. 
From these observations, we conclude that the chiral gravitational effect causes the time variation of the helicity, or the gravitational birefringence.
In what follows we mainly focus on the case where only the left-handed polarization mode exists, 
but the generalization to the situation where both circular polarization modes exist can be done in a straightforward way.

We find a remarkable feature of the chiral gravitational effect. As read from Eq.~\eqref{sol}, the correction in the first term can vanish regardless of the evolution history of $\mu_5$ if the chiral imbalance today is the same as its initial value since it is proportional to the net chemical potential difference between the initial and final values, $\Delta \mu_5$. On the other hand, the second term is in general non-vanishing even if $\Delta{\mu}_5$ eventually vanishes, thanks to the nontrivial phase factor $e^{-2\ri kt}$ in the integrand and depends on the evolution history of $\mu_5$. This implies that the chiral gravitational effect in a time-dependent background leaves the ``memory" of the chiral imbalance in the time evolution history. Since in the early Universe, nontrivial time evolution of $\mu_5$ can be realized %{\it e.g.} 
by chirality-violating processes such as 
the GUT baryogenesis
as well as the chirality flip by the Yukawa interactions, gravitational birefringence may take place in the primordial chiral plasma. In such a situation, the primordial gravitational wave can be a probe to investigate the dynamical chirality-violating processes in the early Universe through the chiral gravitational effect. This property also appears in the expanding background as we will see later. However, one needs to care the possibility that the backreaction to the chiral plasma from gravitational waves might affect the memory of such processes. In the next subsection, we analyze the anomaly equation to quantify the backreaction and show that their effect on the memory of the gravitational waves is negligibly small. 

We should emphasize that the existence of non-zero $\mu_5$ at the beginning does not affect the propagation of gravitational waves through the local contribution represented by the Chern-Simons term. 
This behavior is quite different from the electromagnetic case where the chiral plasma instability is caused by the non-zero chemical potential itself but not its time variation~\cite{Joyce:1997uy,Akamatsu:2013pjd}. 
This different dependence on $\mu_5$ can be ascribed to the spin structure of the field of interest. Technically, the gravitational Chern-Simons term, which involves the spin 2 field, contains more derivatives than that for the spin 1 gauge field. As a result,  $\ddot{\theta} \sim \dot{\mu}_5$ appears in the first order in $\mu_5$. Physically speaking, the structure of the effective action of a spin $n\geq1$ field is determined by the gauge symmetry associated with the field. For graviton, the diffeomorphism determines a possible form of the effective action and as a result, it becomes relevant only when the chemical potential is time-dependent. 
As we quoted in Sec.~\ref{CS}, however, there are non-local contributions to the graviton dispersion relation proportional to $\mu_5^3$ and $T^2\mu_5$~\cite{Manes:2012hf,Jensen:2012kj,Assuncao:2018jkq}, which cannot be treated in our local effective Lagrangian.\footnote{In other words, the absence of the chiral effects in the graviton propagation for $\dot{\mu}_5=0$ would be independent of our perturbative treatment of $\mu_5$ as can be read from Eq.~\eqref{eom_f2}.} 
If such non-local effects dominate the graviton propagation, a constant chemical potential
can modify the gravitational wave dynamics as shown in Ref.~\cite{Sadofyev:2017zqc}.

\subsection{Backreaction to chiral plasma from gravitational waves}\label{br_f}
We have obtained the anomaly corrected gravitational wave solution and found that the time variation of the chiral chemical potential can cause the gravitational birefringence. In order to comprehensively trace the evolution of this system, we must investigate the backreaction from the gravitational waves to the chiral plasma. To our best knowledge, such an issue has not been addressed so far. For this purpose, we here analyze the anomaly equation to investigate the backreaction by using the wave solutions obtained in the previous section. 

Gravitational anomaly expresses the violation of the chiral charge conservation of the fermion due to the presence of non-vanishing topological configuration of the gravitational wave as 
\beq
\nabla_{\mu}j_5^{\mu} = -\frac{1}{12(4\pi)^2}R\tilde{R}. \label{G_anomaly}
\eeq
What physically happens is that the chiral asymmetry of gravitational waves acts as a biased background for the evolution of the Dirac equation~\cite{AlvarezGaume:1983ig}. Let us evaluate the contribution of the right-hand side of the anomaly equation~\eqref{G_anomaly} with the gravitational wave solution obtained in Sec~\ref{GW_f}. In the Minkowski spacetime, the spatial average of $R\tilde{R}$ can be expanded as
\begin{align}
\langle R\tilde{R}\rangle &\simeq 
\partial_t\lnk \epsilon^{ijk} \lmk\dot{h}_{il}\partial_j\dot{h}_{k}^{\ l} -\partial_lh_{im}\partial^{l}\partial_j h_{k}^{\ m}\rmk\rnk \notag \\
&= 4\partial_t \left\{k \left(\left|\dot{h}_{\rm R}(t)\right|^2-\left|\dot{h}_{\rm L}(t)\right|^2\right)-k^3\left(\left|h_{\rm R}(t)\right|^2-\left|h_{\rm L}(t)\right|^2\right)\right\}.\label{RR_f} 
\end{align}
By substituting Eq.\eqref{sol} into Eq.\eqref{RR_f}, we obtain the evolution equation for the chiral charge carried by fermions with the gravitational anomaly 
from the plane wave solution for the left-handed polarization mode, $h_\mathrm{L}^{(0)} (t,z)= A e^{-ik(t-z)}$ or $h_\mathrm{L}^{(0)} (t)= A e^{ikt}, h_\mathrm{R}^{(0)}(t) =0$
as\footnote{If one consider the initially non-helical plane wave as $h_\mathrm{L}^{(0)} (t)= h_\mathrm{R}^{(0)}(t) = A e^{ikt}$, the right-handed polarization mode adds the same contribution as the left-handed polarization mode to the right-hand side of Eq.~\eqref{n5_f}. Thus, for the non-helical initial condition, one should replace $C=1/12(4\pi)^2$ with $\tilde C \equiv 2\times1/12(4\pi)^2$ in Eq.~\eqref{n5_f}.}%the following discussion still holds by re-defining $C \equiv 2\times1/12(4\pi)^2$.}
\begin{align}
\dot{n}_5 %\simeq 
&= C \frac{A^2 k^4}{3 \pi^2 M_\mathrm{Pl}^2} \partial_t \left\{ \int^t_{-\infty}dt'\dot{\mu}_5(t')\cos 2k(t-t')\right\} \left(1+\mathcal{O}\lmk \frac{k \mu_5}{M_{\rm Pl}^2}, \frac{\dot{\mu}_5}{M_\mathrm{Pl}^2}\rmk\right) \notag \\
&= C\frac{A^2k^4}{3\pi^2 M_{\rm Pl}^2}\lnk \dot{\mu}_5(t) -2k\int^t_{-\infty}dt'\dot{\mu}_5(t')\sin2k(t-t')\rnk \left(1+\mathcal{O}\lmk \frac{k \mu_5}{M_{\rm Pl}^2}, \frac{\dot{\mu}_5}{M_\mathrm{Pl}^2}\rmk\right),\label{n5_f}
\end{align}
where $n_5=j_5^0$ and $C \equiv 1/12(4\pi)^2$.
Here we assume that the spatial gradient of the chiral charge is always negligibly small and 
also neglect the contributions from the spatial component of the chiral current. 
%\KK{Once more, the contributions from the plane wave of the right-handed polarization mode is obtained by replacing $\mu_5$ in Eq.~\eqref{n5_f} with $-\mu_5$.} \JK{(JK:This is not correct. Due to the sign flip of $\lambda$, right-handed mode gives the same contribution as in the r.h.s of Eq.~\eqref{n5_f}.)}
From Eq.~\eqref{n5_f}, one can find that the ``modulated'' chemical potential results in the violation of the chiral charge in the fermion sector, which in turn affects the chemical potential. 

When the plasma is thermalized, chiral charge density is related to the chiral chemical potential as $n_5 \simeq \mu_5T^2/6$ for $\mu_5 \ll T$. Assuming that the chirality-changing process is sufficiently slow and the equilibrium is always maintained, the variation of the charge can be written in terms of the chemical potential. As a simple example, let us first consider the case where the non-zero constant $\mu_5$ exists but there are no other dynamical chiral charge violating processes. In this case, the evolution of $\mu_5$ is determined merely by the anomaly induced backreaction (Eq.~\eqref{n5_f}). Then we can write the kinetic equation of $\mu_5$ in the following form:
\begin{align}
\dot{\mu}_5(t) &= \delta  \ \partial_t \left\{  \int^t_{-\infty}dt'\dot{\mu}_5(t')\cos2k(t-t')  \right\}\notag \\
&= %\sout{\frac{2C}{\pi^2}A^2\frac{k^4}{T^2M_{\rm Pl}^2}}
\delta\lnk \dot{\mu}_5(t) -2k\int^t_{-\infty} dt'\dot{\mu}_5(t')\sin2k(t-t')\rnk,\label{mu_kin}
\end{align}
where $\delta \equiv \frac{2C}{\pi^2}A^2\frac{k^4}{T^2M_{\rm Pl}^2}$ which characterizes the strength of the backreaction. By substituting the integrated form of Eq.\eqref{mu_kin} into its differentiated form, one can obtain the following second-order differential equation, 
\beq
(1 - \delta)\Delta\ddot{\mu}_5 + 4k^2\Delta\mu_5 = 0. \label{no_source}
\eeq
Hereafter we consider the case $\delta \ll 1$, which is realized with sufficiently small $A$ 
and $k$.\footnote{We however assume that $k$ is not too small so that 
the conditions for the non-local effects
to be negligibly small, $\mu_5, T \ll k$, are satisfied.}
Clearly this equation describes the harmonic oscillator. However, Eq.~\eqref{mu_kin} gives the initial condition $\Delta\dot{\mu}_5 = 0$ which results in the trivial solution $\Delta\mu_5(t) = 0$. As we have emphasized in the previous section, this behavior is completely different from the CME where the instability is triggered just by the existence of the non-zero chemical potential through the local anomalous contribution.\footnote{This point also shows the difference between CME and the chiral gravitational effect, as the constant chemical potential causes non-trivial evolution of the system for the former through the local Lagrangian term but for the latter only through the non-local contributions.}

Next, we consider the case where the non-trivial time evolution of $\mu_5$ is sourced by a chirality-inducing process, which plays the role of an external force to the static harmonic oscillator.
The backreaction of the chiral gravitational effect~\eqref{n5_f} is expected to perturb this dynamics slightly. In this situation, we can expand the chiral chemical potential in the following way:
\beq
\mu_5(t) = \mu_5^{(0)}(t)+\mu_5^{(1)}(t), 
\eeq
where the first term represents the chirality generation governed by the dominant process and the second term describes the perturbative backreaction due to the anomaly equation~(Eq.~\eqref{n5_f}). 
Having in mind the decay of heavy particles, such as 
heavy colored Higgs field $H^c \subset  {\bm 5}$ in the case of SU(5) GUT baryogenesis~\cite{Kamada:2018tcs,Domcke:2020quw}, we consider a model where $\mu_5^{(0)}(t)$ is given as
\beq
\dot{\mu}_5^{(0)}(t) = 
 \left\{ \begin{array}{ll} 0, & \text{for} \quad t<0, \\ 
\Gamma\exp(-t/\tau), & \text{for} \quad t\geq 0,
\end{array}\right. \label{source}
\eeq
where the decay of the heavy particle turns on at $t=0$ 
with $\tau$ being the lifetime of the heavy particle.
$\mu_5^{(1)}(t)$ then can be evaluated perturbatively  in powers of the parameter~$\delta$ which characterizes the strength of the backreaction as
\beq
\dot{\mu}_5^{(1)}(t) = \delta\lnk \dot{\mu}^{(0)}_5(t) -2k\int^t_0dt'\dot{\mu}_5^{(0)}(t')\sin2k(t-t')\rnk
+\mathcal{O}(\delta^2),
%-\Gamma\mu_5^{(1)}(t).
\label{mu_kin_pert}
\eeq
where the right-hand side represents the contribution from the anomaly as in Eq.~\eqref{mu_kin}.
%With the initial condition $\Delta\mu_5(0) = 0$ and eq.~\eqref{source} and~\eqref{mu_kin_pert}, 
Using Eqs.~\eqref{source} and~\eqref{mu_kin_pert} with the initial condition $\Delta\mu_5(0) = 0$, $\mu_5(t)$ 
 %up to the first order in $\delta$ 
 can be expanded as
\begin{align}
\Delta\mu_5(t) &= \int_{-\infty}^t dt'  ( {\dot \mu}_5^{(0)}(t')+{\dot \mu}_5^{(1)}(t') ) \notag \\
&\simeq  \Gamma\tau \lmk 1-e^{-\frac{t}{\tau}}\rmk - \frac{\Gamma\tau\delta}{1+4k^2\tau^2}\lmk\cos(2kt)+2k\tau\sin(2kt)-e^{-\frac{t}{\tau}}\rmk +\mathcal{O}(\delta^2).\label{linear_osc} 
\end{align}
The first term is nothing but the leading contribution from the source term whereas the second term represents the backreaction from %the 
gravitational waves due to the gravitational anomaly. The latter is significantly suppressed at least by a factor of $\delta$, but remains even in the late time where the source contribution eventually disappears. 
Compared to the previous discussion, here a non-zero initial ${\dot \mu}_5$ is allowed 
with the help of the non-vanishing source term at an earlier time. Therefore a non-zero oscillatory term, which is a solution to Eq.~\eqref{no_source} with $\delta = 0$, appears as the consequence of the backreaction. Thus, one can conclude that the oscillatory conversion between the chirality of the plasma and the helicity of the gravitational waves is triggered by the source.

By investigating the structure of Eq.~\eqref{linear_osc} in more depth,
we can read off the following general features which appear when one incorporates the external chirality-changing process into the system, which eventually disappears at a late time. At an early time, the backreaction (${\dot \mu}_5^{(1)}$) is suppressed by $\delta$ and the dynamics of $\mu_5(t)$ is determined by the source. As discussed in the previous subsection, 
${\dot \mu}_5^{(1)}$ contains the contribution from
the anomaly induced correction of the gravitational waves in Eq.~\eqref{sol}, which 
includes the
%contains 
time integration of $\dot{\mu}_5e^{-2\ri k t}$.  
This is what we call the ``memory effect''. % \KK{\sout{and does not decay with time}}. 
If the magnitude of $\dot{\mu}_5^{(0)}$ monotonically decays as in this model, the correction to the gravitational wave propagation is dominated by the ``memory'' of the source dynamics in early times.
Then eventually the kinetic equation for the chiral chemical potential is 
dominated by the contributions from the memory effect. 
In the right hand side of Eq.~\eqref{mu_kin_pert}, for example, it corresponds to 
the contribution from the second term in the parentheses. By performing the integral over the early times, the memory contribution can be approximated as 
$
\propto \delta k C \sin (2k t + \alpha). 
$
% with 
The factor $C$ denotes the result of the integral and is dominated by the early time contribution as 
$C\sim {\dot \mu}_\mathrm{ini} \Delta t$, where $\Delta t$ is the time scale of the decay of the source. One can see that this contribution remains as the oscillating source even in the late time. Thus we can understand that the non-decaying oscillation in the second term of Eq.~\eqref{linear_osc} originates from this early memory contributions.\footnote{For much later time, $t \gg \Delta t/\delta$, the integrated late time contributions of the source might become comparable to the early time contributions in general.}
Nevertheless, when the source contribution becomes comparable to the backreaction, we can no longer treat the source $\dot{\mu}_5^{(0)}$ and backreaction $\dot{\mu}_5^{(1)}$ separately.
This means that there should be a non-trivial interplay between source and the backreaction in much later time. 
%As for the chirality inducing process, however, such a modification changes the dynamics only at higher order in $\delta$, and hence we expect that the prediction given in Eq.~\eqref{linear_osc} is %considered to be valid.
Does such a regime affect the gravitational wave behavior? The answer is
negative. Although the interplay between the source and backreaction in the late time can be non-trivial, their amplitude is expected to be suppressed at least by a factor of $\delta$. Since the memory, or the helicity accumulated in gravitational waves, consists of the dominant source contribution in the early time, such a late time contribution should be negligible for the modifications of the gravitational wave propagation.
From these observations, we expect that in general the leading contribution in the backreaction comes from the early time memory effect. However, the late time chemical potential would be ``sourced''  by the early time memory effect, which may lead to nontrivial dynamics of the chemical potential.

\begin{figure}[htbp]
  \centering
%  \subfloat[]{
  \includegraphics[width=10cm,clip]{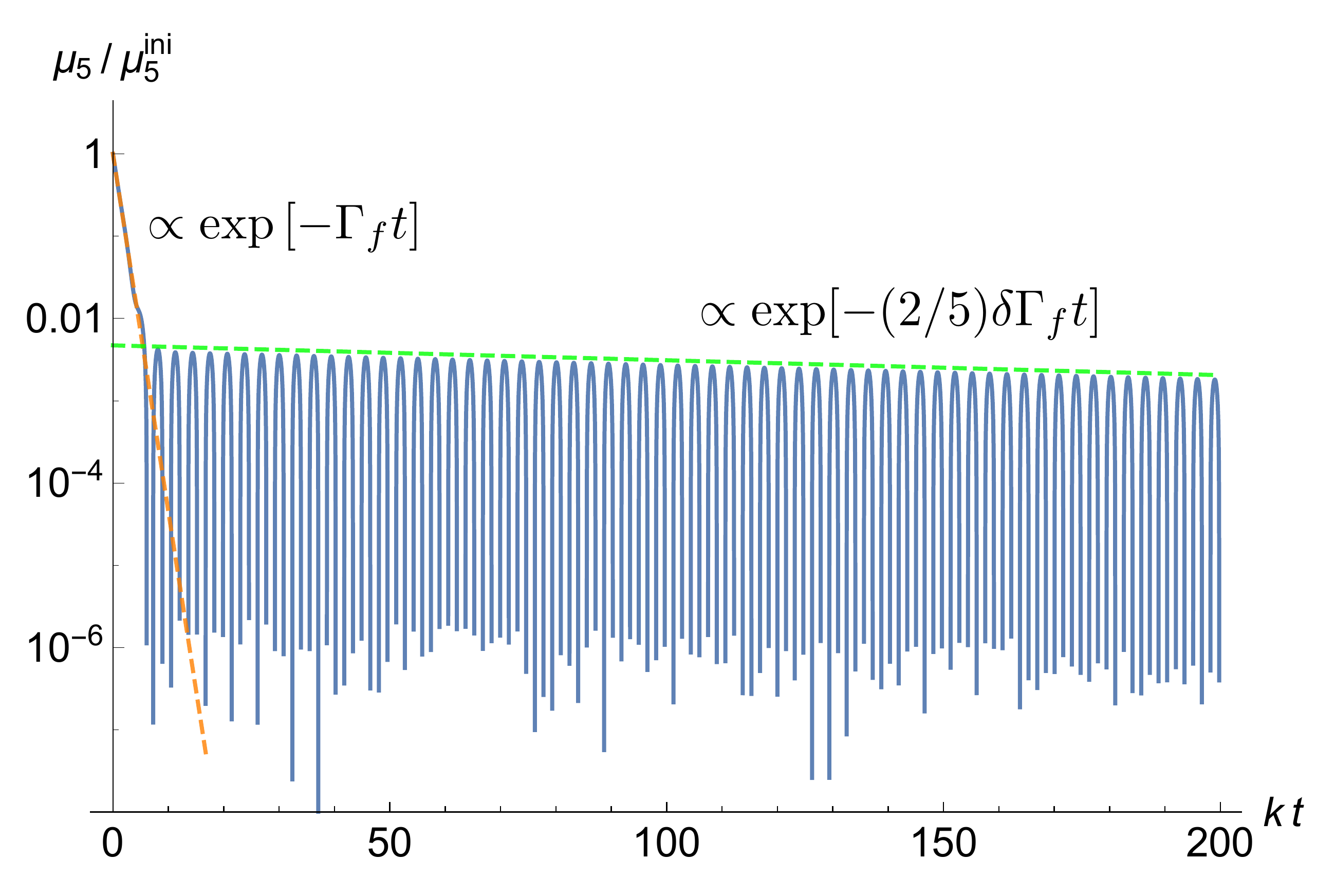}
%  \subfloat[]{
%  \includegraphics[width=7.5cm,clip]{osc2.pdf}}
    \caption{Numerical calculation of the solution to Eq.~\eqref{third} in the earlier times
    with $\Gamma_f = k$ and $\delta = 0.01$. %In order to clarify how non-zero but small backreaction affects the system, we take $\Gamma_f = k = \mu_5^{\rm ini} = 1$ and $\delta = 0.01$ here. 
    For $t \lesssim 1/\Gamma_f$, the initial asymmetry is washout in a rate proportional to 
    $\exp[-\Gamma_f t]$. For $t \gg 1/\Gamma_f$, the backreaction dominates 
    and the chiral chemical potential shows the slowly decaying oscillation in proportion to 
    $\exp[-c \delta \Gamma_f t]$ with $c \simeq 2/5$ in this parameter set. }
%    As one can see, after the exponential decay in the earliest time, the system dominated by the oscillation arising from the backreaction and it is eventually damped. However, it seems hard to obtain precise prediction for the late time behavior due to the numerical errors.}
    \label{fig2}
\end{figure}

To show the generality of the discussion in the above, let us investigate the effect of 
another example, the chirality-flipping interactions, which is responsible for the decay of the chiral asymmetry.
Note that the chirality is not an exact symmetry of the Standard Model of particle physics
and is flipped by Yukawa interactions.
To mimic such chirality-flipping interactions, we consider the following term, 
\begin{equation}
{\dot \mu}_5(t) \ni  -\Gamma_f \mu_5(t), \quad \text{for} \quad t\geq 0, \label{flip}
\end{equation}
with non-vanishing initial chiral asymmetry, $\mu_5^\mathrm{ini} \not = 0$ at $t=0$. 
Here $\Gamma_f$ is the chirality-flipping rate, which turns on at $t=0$. 
The anomaly-corrected kinetic equation 
for the chiral chemical potential is then given by
\begin{equation}
{\dot \mu}_5(t) = - \Gamma_f \mu_5(t)  + \delta\lnk \dot{\mu}_5(t) -2k\int^t_{0}
 dt'\dot{\mu}_5(t')\sin2k(t-t')\rnk. \label{mu_kin_flip} 
\end{equation}
At an early time where the backreaction can be treated perturbatively, the dynamics is dominated by the chirality-flipping effect and $\mu_5(t)\sim \mu_5^{\rm ini}e^{-\Gamma_f t}$, 
similar to the chirality inducing process. 
As discussed in the above for that 
case, we expect that the dominant memory effect on gravitational waves originates from this early stage. 
In the later time, 
the chemical potential exponentially decays and eventually the backreaction due to the early memory effect becomes comparable to the leading contribution. Then, one needs to solve Eq.~\eqref{mu_kin_flip} without a perturbative method in $\delta$. 
Practically, it is useful to replace the integro-differential equation~\eqref{mu_kin_flip} with a differential equation. 
In a similar way how we obtained Eq.~\eqref{no_source}, 
Eq.~\eqref{mu_kin_flip} can be rewritten as 
\begin{equation}
(1-\delta) \dddot{\mu}_5(t) + \Gamma_f \ddot{\mu}_5(t) + 4 k^2 \dot{\mu}_5(t) + 4 k^2 \Gamma_f \mu_5(t) = 0.\label{third}
\end{equation}
Note that the initial conditions for $\mu_5$ and its time derivatives at $t=0$ are fixed by Eq.~\eqref{mu_kin_flip} and its differentiated form up to the second order derivative, namely, $\mu_5(0) = \mu_5^{\rm ini}$, $\dot{\mu}_5(0) \simeq -\Gamma\mu_5^{\rm ini}(1+\delta)$, $\ddot{\mu}_5(0) \simeq \Gamma^2\mu_5^{\rm ini}(1+2\delta)$. 
Here we solve Eq.~\eqref{third} numerically for the parameters
$\Gamma_f/k = 1$, and $\delta= 0.01$. 
Fig.~\ref{fig2} shows the evolution of the chiral asymmetry in earlier times.
We can see that the initial asymmetry is exponentially damped in proportion to $\exp[-\Gamma t]$ and a small slowly-decaying oscillation follows with $\propto \exp[- (2/5) \delta \Gamma t]$. 
This behavior can be understood as follows.
By taking $\delta = 0$, the formally zero-th order solution in $\delta$ can be obtained as 
\begin{equation}
\mu_5(t) = \mu_5^{\mathrm{I}} \exp\left[-\Gamma_f t\right] + \mu_5^\mathrm{II} \cos (2 k t) +\mu_5^\mathrm{III} \sin (2 k t). \label{third_sol}
\end{equation}
While the first term represents the dominant exponentially damping contribution,
the second and third contributions  
appear due to the backreaction and show the oscillatory behavior with the same frequency to the one in Eq.~\eqref{linear_osc}. 
By considering the connection to the initial condition, one can see that $\mu_5^\mathrm{II}$ and $\mu_5^\mathrm{III}$ is the order of $\delta$. 
As a result, %we can conclude that 
$\mu_5$ shows an exponential decay and a small oscillation around it in the early time, which is similar to the chirality inducing process. 
One can see that the oscillatory part decays slowly, which is due to the finite $\delta$ correction to Eq.~\eqref{third_sol} from Eq.~\eqref{third}. Such a correction should be in proportion to $\exp[- c \delta \Gamma t]$ so that the decay disappears if one takes $\delta \rightarrow 0$ limit. Here $c$ is a numerical factor, which found to be 2/5 in this parameter choice.
Note that such decay rates can be obtained by solving the characteristic equation of ~\eqref{third}, especially for the specific limit of $\Gamma/k$.
We would like to emphasize that this is consistent with the discussion in the above. 
Here the parenthesis in the kinetic equation~\eqref{mu_kin_flip}, or the backreaction from the gravitational wave propagation, 
dominated by the memory effect in the early time, $\sim \delta k C_f \sin (2k t + \alpha)$, now acts as the source term and is equilibrated to the chirality-flipping term. 
Here $C_f$ denotes the dominant contribution of the integral over the early times, and 
$\alpha$ denotes the phase shift due to the time-dependent ${\dot \mu}_5$. 
This leads to the slowly decaying oscillation of $\mu_5$.

As a concluding remark of this section, let us re-emphasize the memory effect on gravitational waves.
As mentioned in the previous subsection, gravitational birefringence takes place according to the history of the evolution of $\dot{\mu}_5$ during the propagation. In both the chirality induction and flipping processes, we have seen that the dynamics of $\mu_5$ is determined by the source at an early time and the backreaction is generically negligible there. The dominant memory effect originates from this early stage. However, in a reasonable setup, eventually the external source 
for $\mu_5$ decays and becomes comparable to the backreaction coming from the early time memory effect. The interplay between various contributions to $\mu_5$ dynamics makes its behavior rather complicated, but eventually $\mu_5$ would behave in an oscillatory form forced by the oscillating source due to the backreaction
at least for a duration with a factor of $1/\delta$. Note that the much later time behavior of the chemical potential, which is rather non-trivial, would have less significant magnitude than that of the early time, which is dominated by the external sources.     
Thus, we conclude that the gravitational waves memorize the dominant evolution of the chirality 
and the memory is hardly affected by the backreaction.
It is quite interesting that the memory effect induced by dynamics of $\mu_5$ in early times affects (or dominates) the late time behavior of the chemical potential itself through the backreaction.
As we will see in the next section, such a property of the memory effect does not change even in the FRW background case where the cosmological expansion also contributes to the birefringence through the chiral gravitational effect.

\section{Chiral gravitational effect in the expanding background}\label{br}
As %we 
mentioned in Sec.~\ref{CS}, the non-conformal nature of the graviton gives a non-trivial background spacetime dependence on the propagation of the gravitational waves through the chiral gravitational effect. Having in mind the application to the cosmology, we discuss the chiral gravitational effect in the FRW spacetime in this section. We will present how the cosmic expansion %affects 
plays a role in the chiral gravitational effect as follows. We first calculate the wave solutions for the modified equation of motion in a general time-dependent chiral asymmetric fermion background in Sec.~\ref{GW_FRW}. 
Then, we apply them to the case for the chiral chemical potential diluted by the cosmic expansion
and evaluate the backreaction on the chiral chemical potential in Sec.~\ref{br_FRW}. 

\subsection{Propagation of gravitational waves}\label{GW_FRW}
Here we discuss the anomaly corrected gravitational wave solution in the expanding Universe and demonstrate how the effect of the expanding background appears in the chiral gravitational effect. 
Let us consider the flat FRW background $ds^2 = a(\eta)^2[-d\eta^2+d{\bm x}^2]$ where $\eta$ represents the conformal time. With the help of the covariant expression, we can again start from the effective action obtained in Sec.~\ref{CS}, namely ${S}_{\text{E-H}}+ { S}_\mathrm{eff}^{\rm PV} = \int d^4 x \sqrt{-g} [(M_\mathrm{Pl}^2/2)R-\theta R {\tilde R} ]$.
Note that the spurious scalar field should now be related to the chiral chemical potential as $\partial_{\eta}\theta = (\mu_5a(\eta)/{192\pi^2},0,0,0)$
since $\partial_\mu \theta$ acts as a conformally-invariant field, which should be constant
in the absence of the chirality-changing interactions.

In the Fourier space, one can expand metric perturbation in the circularly polarized basis as
\beq
h_{ij}(\eta,\textbf{x}) = \frac{1}{(2\pi)^3}\int d^3k\sum_{A = \mathrm{R,L}}p_{ij}^{A} (\bm{k})h_{\bm{k}}^{A} (\eta)e^{i\bm{k}\cdot \bm{x}}, \label{Fourier}
\eeq
where the circular polarization tensors in the three-dimensional momentum space\footnote{Here we have generalized the circular polarization tensor, which was introduced in 
Sec.~\ref{GW_f} for ${\bm k} \propto (0,0,1)$.} satisfy the following equations,
\begin{equation}
\begin{split}
  p_{ij}^\mathrm{R}({\bm k})p^{ij\mathrm{R}}({\bm k}) &= p^\mathrm{L}_{ij}({\bm k})p^{ij\mathrm{L}}({\bm k}) = 0,\\
  p_{ij}^\mathrm{R}({\bm k})p^{ij\mathrm{L}}({\bm k}) &= 2,\\
  k_p\epsilon^{mpj} p^A_{ij} ({\bm k}) &= -i\lambda^A_{\bm k} k \ p^{m \  A}_{\ \ i}({\bm k}) \ (\text{for} \ A=\mathrm{L, R}), 
\end{split} \label{cptr_k}
\end{equation}
with $\lambda^\mathrm{R}_{\bm k} = +1, \lambda^\mathrm{L}_{\bm k} = -1$.\footnote{Note that $\lambda_{\bm {k}}^A$ changes its sign as $\lambda^\mathrm{R}_{-\bm k} = -\lambda^\mathrm{R}_{\bm k} = -1, \lambda^\mathrm{L}_{-\bm k} = -\lambda^\mathrm{L}_{\bm k} = +1$, with respect to the reflection of the wave vector.} The polarization tensors satisfy
$p^A_{ij}({\bm k}) = p^A_{ij}(-{\bm k}) = (p^{\bar A}_{ij}({\bm k}))^*\ (A = \mathrm{L,R}; \bar{\mathrm{L}} = \mathrm{R}, \bar{\mathrm{R}} = \mathrm{L})$
so that the graviton satisfies the reality condition with $h_{{\bm k}}^\mathrm{R} = (h_{-{\bm k}}^\mathrm{L})^*$.
We find the quadratic action of the tensor perturbation around the FRW background in the momentum space as
\beq
 {\cal S}_{\text{E-H}}+{\cal S}_{\rm eff} \simeq \frac{M_{\rm Pl}^2}{4} \int d\eta d^3k \sum_{A = L,R} a^2(\eta)\left[1 + \lambda^A_{\bm k} \frac{k}{24\pi^2M_{\rm Pl}^2}\frac{\mu_5}{a(\eta)}\right](|(h^A_{{\bm k}})'|^2 - k^2|h^A_{{\bm k}}|^2),
 \eeq
where the prime denotes the derivative with respect to the conformal time $\eta$. 
Requiring the absence of ghost-like and strong-coupling modes below the Planck scale for the physical momentum $k/a(\eta)$, we obtain the same upper bound on the value of chiral chemical potential as Eq.~\eqref{constraint}.

We are interested in the early epoch where the primordial gravitational waves are generated by a dynamical mechanism and
the Universe eventually consists of chiral asymmetric thermal plasma. %Hence w
We here investigate the gravitational wave %generation or
propagation in the radiation-dominated Universe without any entropy production,\footnote{See Ref.~\cite{Barrie:2017mmr} for the calculation of matter-dominated era.} in which the scale factor evolves with time as $a(\eta) = a_0(\eta/\eta_0)$ and the temperature of the Universe dilutes roughly as $T\propto a^{-1}(\eta)$. 
We take $\eta=\eta_0$ to be the ``initial'' time and does not take into account the process
to generate the initial asymmetry. Well after the generation of gravitational waves and the source term has disappeared, the mode equation is written as
\beq
\begin{split}
  (h_{\bm k}^A)^{\prime\prime} +\frac{2}{\eta}(h_{\bm k}^A)^{\prime}+ k^2h_{\bm k}^A &= -\frac{\lambda^A_{\bm k}k}{24\pi^2M_{\rm Pl}^2}\lnk\frac{\mu_5\eta_0}{a_0\eta}\lmk\frac{d^2}{d\eta^2}+\frac{2}{\eta}\frac{d}{d\eta}+ k^2\rmk+\lmk\frac{\mu_5\eta_0}{a_0\eta}\rmk^{\prime}\frac{d}{d\eta}\rnk h_{\bm k}^A\\
  &\equiv -\frac{\lambda^A_{\bm k}k}{24\pi^2M_{\rm Pl}^2} \hat{F}_k(\eta)h_{\bm k}^A. \label{eom_FRW}
\end{split}
\eeq
Once more, we perturbatively solve the equation of motion using the Green function method. In the FRW background, the Green function is given by
\beq
G(\eta,\tilde{\eta}) = \Theta(\eta -\tilde{\eta})\tilde{\eta}\frac{\sin k(\eta-\tilde{\eta})}{k\eta},
\eeq
which satisfies $(\partial_\eta^2+(2/\eta)\partial_\eta+k^2)G(\eta,\tilde{\eta})=\delta(\eta-\tilde{\eta})$.
From the zero-th order solution of Eq.~\eqref{eom_FRW} with $\mu_5={\dot \mu}_5 = 0$,\footnote{Since the other solution $h_{\bm k}(\eta)\propto \cos k\eta/(k\eta)$ is a decaying mode at the superhorizon scales, $k \eta\ll 1$, we do not consider it here.} which is given by 
\beq
h_{\bm k}^{A,(0)}(\eta) = h_{\rm in}^{A}({\bm k})\frac{\sin k\eta}{k\eta}, 
\eeq
we can evaluate the first order correction as
\beq
\begin{split}
  h_{\bm k}^{A,(1)}(\eta) - h_{\bm k}^{A,(0)}(\eta) &= -\frac{\lambda^A_{\bm k}k}{24\pi^2M_{\rm Pl}^2}\int_{\eta_0}^{\infty}d\tilde{\eta}G(\eta,\tilde{\eta})\hat{F}_k(\tilde{\eta})h_{\bm k}^{A,(0)}\\
  &= -\frac{\lambda^A_{\bm k}kh_{\rm in}^{A}({\bm k})}{24\pi^2M_{\rm Pl}^2}\int_{\eta_0}^{\eta}d\tilde{\eta}\lmk\frac{\mu_5\eta_0}{a_0\tilde{\eta}}\rmk^{\prime} \frac{\sin k(\eta-\tilde{\eta})}{k\eta}\lmk\cos k\tilde{\eta}-\frac{\sin k\tilde{\eta}}{k\tilde{\eta}}\rmk.\label{sol_FRW}
\end{split} 
\eeq
Compared to the Minkowski background case (Eq.~\eqref{sol}), not only the intrinsic variation of the chemical potential but also the cosmological expansion affects the propagation of gravitational waves through the term $(\mu_5/a(\eta))^{\prime}$. 
This means that the fermion background gives the non-zero correction to the gravitational waves even when the comoving chiral asymmetry is conserved and $\mu_5$ just decays as $\mu_5 \propto a^{-1} \propto \eta^{-1}$ according to the cosmic expansion.
Such a background curvature dependence is ascribed to the non-conformal nature of the graviton and is obviously different from the CME where one can drop the scale factor dependence by moving to the comoving frame. 
We would like to emphasize that we need to take into account the effect of the cosmic expansion in order to quantify the memory effect of gravitational waves, which we have discussed in the Minkowski spacetime case. 
With this solution, we investigate how the cosmological expansion plays a role in the interplay between the chiral gravitational waves and chemical potential through the gravitational anomaly in Sec.~\ref{br_FRW}.

\subsection{The effect of the diluting chemical potential}\label{br_FRW}
As we explained in Sec.~\ref{GW_FRW}, the cosmic expansion gives a non-trivial contribution to the evolution of the gravitational waves through the
chiral gravitational effect, which is not the case in the CME where all the fields involved are conformal. In order to quantify the effect of the cosmic expansion, we investigate the behavior of the chiral plasma and gravitational waves in the expanding background
in the absence of the chirality-changing interactions. 
As is in the Minkowski spacetime case, let us consider a plasma with a chiral chemical potential $\mu_5 (\ll T)$ or $n_5 \simeq \mu_5 T^2/6$. Note that the temperature scales as $T \propto a^{-1}(\eta)$ in the expanding background. We divide them into the background that is simply diluted by the cosmic expansion and the modulation due to the backreaction as
\beq
\mu_5(\eta) \simeq \frac{a_0}{a(\eta)}\lmk\mu_5^{(0)} + \mu_5^{(1)}(\eta)\rmk. %+ \mathcal{O}(\delta^2),
\eeq
Here $\mu_5^{(0)}$ is a constant that represents the initial chiral asymmetry of the system
whereas the modulation due to the backreaction $\mu_5^{(1)}$ 
is expected to be sufficiently small. 
The leading correction to the gravitational waves caused by the cosmological dilution of the chemical potential is obtained by substituting $\mu_5\simeq\mu_5^{(0)} (a_0/a(\eta))$ into Eq.~\eqref{sol_FRW}:
\beq
\begin{split}
h_{\bm k}^{A,(1)}(\eta) &=  h_{\rm in}^{A}({\bm k})\lnk\frac{\sin k\eta}{k\eta}-\frac{\lambda^A_{\bm k}k}{24\pi^2M_{\rm Pl}^2}\int_{\eta_0}^{\eta}d\tilde{\eta}\lmk\frac{a_0 \mu_5^{(0)}}{a(\tilde{\eta})^2}\rmk^{\prime} \frac{\sin k(\eta-\tilde{\eta})}{k\eta}\lmk\cos k\tilde{\eta}-\frac{\sin k\tilde{\eta}}{k\tilde{\eta}}\rmk\rnk\\
&=h_{\rm in}^{A}({\bm k})\lnk\frac{\sin k\eta}{k\eta}-\frac{\lambda^A_{\bm k}k\mu_5^{(0)}}{24\pi^2M_{\rm Pl}^2a_0}\int_{\eta_0}^{\eta}d\tilde{\eta}\lmk\frac{\eta_0^2}{\tilde{\eta}^2}\rmk^{\prime} \frac{\sin k(\eta-\tilde{\eta})}{k\eta}\lmk\cos k\tilde{\eta}-\frac{\sin k\tilde{\eta}}{k\tilde{\eta}}\rmk\rnk\\
&\equiv h_{\rm in}^{A}({\bm k})\lnk\frac{\sin k\eta}{k\eta}-\frac{\lambda^A_{\bm k}k \mu_5^{(0)}}{24\pi^2M_{\rm Pl}^2a_0}f(k, \eta)\rnk,
\end{split}
\eeq
where we have defined $f(k,\eta)$ for the notational convenience. 
Here the first order correction $\delta h_{\bf k}^A (\eta) \equiv h_{\bm k}^{A,(1)} (\eta) - h_{\bm k}^{A,(0)} (\eta) = \frac{\lambda^A_{\bm k}k \mu_5^{(0)}}{24\pi^2M_{\rm Pl}^2a_0}f(k,\eta) h_{\rm in}^{A}({\bm k}) $ is the linear order in $\mu_5^{(0)}$. 
Since apparently $f(k, \eta)$ is nonzero in general and $\lambda_{\bm k}^A$ has opposite signs 
for the left- and right-polarization modes, we conclude that the dilution of the chiral chemical 
potential due to the cosmic expansion also causes the gravitational birefringence as expected.

Before investigating the properties of the correction to gravitational wave represented by~$f(\eta)$, we need to confirm that the backreaction is sufficiently small, which guarantees the validity of the treatment in the above. Let us investigate the anomaly equation to quantify the backreaction to the chiral asymmetry carried by the fermions. The modified gravitational waves %in turn 
contribute to the evolution of the chiral chemical potential through the gravitational anomaly. With the perturbative expansion, 
the anomaly equation is rewritten as
\beq
\partial_{\eta}(\mu_5^{(1)}) \simeq -\frac{6}{a_0a^2T^2}\frac{1}{12(4\pi)^2}a^4\langle R\tilde{R}\rangle,\label{kin_FRW}
\eeq
where we have used $aT = \mathrm{const}.$ Note that $\langle R\tilde{R}\rangle$ does not depend on $\mu_5^{(1)}$ but on $\mu_5^{(0)}$ in our perturbative treatment. 
Here the Chern-Pontryagin density can be expanded as
\beq
a^4\la R\tilde{R}\ra = \frac{2}{V}\partial_{\eta}\lkk\int \frac{d^3k}{(2\pi)^3}\lnk  
k\left(|(h^\mathrm{R,(1)}_{{\bf k}}(\eta))'|^2 - |(h^\mathrm{L,(1)}_{{\bf k}}(\eta))'|^2\right)
- k^3\left(|h^\mathrm{R,(1)}_{{\bf k}}(\eta)|^2-|h^\mathrm{L,(1)}_{{\bf k}}(\eta)|^2\right)\rnk\rkk, 
\eeq
where $V \equiv \int d^3 x$ is the spatial volume of the system. In the case where the chirality imbalance %in the 
of gravitational waves is absent in the initial distribution, namely, $|h_{\rm in}^L({\bf k})|^2 = |h_{\rm in}^R({\bf k})|^2=|h_0({\bf k})|^2$, 
the leading order terms cancel each other and the term linear in $\mu_5^{(0)}$ becomes the lowest contribution as
\beq
    a^4\la R\tilde{R}\ra =  \frac{2}{V}\partial_\eta\Biggl[\int\frac{d^3k}{(2\pi)^3}\frac{k^3 \mu_5^{(0)}|h_0({\bf k})|^2}{6\pi^2M_{\rm Pl}^2a_0}\left\{\frac{\sin k\eta}{\eta}f(k,\eta)-\left(\frac{\cos k\eta}{k\eta}-\frac{\sin k\eta}{k^2\eta^2}\right)f'(k,\eta)\right\}+\cdots\Biggr].  \label{rrdualfrw}
\eeq
Here the ellipses denote the higher order terms in $\delta h_{\bf k}^A$ and hence of $\mathcal{O}((\mu_5^{(0)})^2)$. 

For simplicity, let us consider the case where the initial spectrum is localized in a specific scale $k_0$ as $|h_0({\bf k})|^2 = A^2V\delta({\bm k} - {\bm k}_0)$, which corresponds to the case we discussed in the Minkowski spacetime~(Sec.~\ref{GW_f}). %Then 
The $k$-integration in Eq.~\eqref{rrdualfrw} can be performed as
\beq
a^4\la R\tilde{R}\ra =  2\partial_\eta\lkk\frac{k^3_0 \mu_5^{(0)} A^2}{48 \pi^5 M_{\rm Pl}^2a_ 0}\lnk\frac{\sin k_0\eta}{\eta}f(k_0,\eta)-\left(\frac{\cos k_0\eta}{k_0\eta}-\frac{\sin k_0\eta}{k_0^2\eta^2}\right)f'(k_0,\eta)\rnk\rkk+\mathcal{O}((\mu_5^{(0)})^2).
\eeq
Now, by comparing this with Eq.~\eqref{kin_FRW}, the formal expression of the first order correction to the chemical potential is given by
\beq
\mu_5^{(1)}(\eta) \simeq -\frac{A^2\mu_5^{(0)}}
{6(2\pi)^7}\frac{k_0^2}{a^2T^2}\frac{k_0^2}{a^2_0M^2_{\rm Pl}}\lnk\frac{\sin k_0\eta}{k_0 \eta}f(k_0,\eta)-\left(\frac{\cos k_0\eta}{k_0\eta}-\frac{\sin k_0\eta}{k_0^2\eta^2}\right)\frac{f'(k_0,\eta)}{k_0}\rnk+ C_0,\label{correction}
\eeq
where $C_0$ is the integration constant to make $\mu_5^{(1)}(\eta_0) = 0$. As expected, there appears a prefactor $A^2k_0^4/a_0^2 a^2T^2M_{\rm Pl}^2$, which is similar to the factor $\delta$ in Eq.~\eqref{mu_kin}. %, appears. 
This is smaller than the unity if $A$ and $k_0$ are sufficiently small as in the Mikowski spacetime case.\footnote{Once more, 
$k_0$ should not be too small so that the non-local effects can be neglected.} 
Let us assume that $k_0$ lies in the
superhorizon scale at $\eta_0$, $k_0 \eta_0 \ll 1$, and then re-enter the horizon during the radiation-dominated era. In this case, the late time behavior of $f(k_0,\eta)$ can be expressed as
\beq
f(k_0,\eta) \simeq -\frac{k_0\eta_0}{9}\frac{\eta_0}{\eta}\lnk3\pi\cos(k_0\eta)+(6\gamma-5+6\log(2k_0\eta_0))\sin(k_0\eta)\rnk \left(1+ \mathcal{O}\lmk\frac{1}{k_0^2\eta^2}\rmk\right),\label{fff}
\eeq
where $\gamma$ is the Euler-Mascheroni constant. 
From these equations, we can read off the damped oscillating behavior of $\mu_5^{(1)}$,  
whose amplitude decays as $\mu_5^{(1)}\propto \eta^{-2} \propto a(\eta)^{-2}$. This implies that the cosmological dilution of chiral chemical potential actually causes the continual conversion between the chirality carried by fermions and the helicity of the gravitational waves as we observed in the Minkowski spacetime  in Sec.~\ref{br_f}. However, the associated oscillation of the chiral chemical potential decays much faster than the dilution itself. %, \KK{which means that the memory effect in the chiral chemical potential is weaker than that in the Minkowski spacetime. Nevertheless, 
In addition to the smallness of the prefactor $A^2k_0^4/a_0^2 a^2T^2M_{\rm Pl}^2$, 
this rapid decay ensures that the backreaction can safely be neglected in the propagation of the gravitational waves under the diluting chiral chemical potential in the FRW spacetime. This behavior is in contrast to that in the models in the Minkowski spacetime where the backreaction eventually dominates the dynamics of chiral chemical potential. 

Now we have confirmed the smallness of the backreaction, 
before closing let us investigate the propagation of the gravitational waves under the expanding background in more detail. From Eq.~\eqref{fff}, the late time behavior of the corrected gravitational wave is explicitly given by
\begin{equation}
   h^{L,R}(\eta) \sim \frac{A}{k_0\eta}\left(\sin k_0\eta\pm\frac{k_0^3\eta_0^2\mu_5^{(0)}}{216\pi^2M_{\rm Pl}^2a_0^2}\lnk3\pi\cos(k_0\eta)+(6\gamma-5+6\log(2k_0\eta_0))\sin(k_0\eta)\rnk\right),
\end{equation}
for the superhorizon mode at $\eta_0$, $k_0\eta_0 \ll 1$,
where the sign $+(-)$ is for the left (right) mode.
The first term in the parentheses corresponds to the injected gravitational wave and the second term is the modulation caused by the chiral gravitational effect. Since both have the same scaling law, $\propto 1/\eta$, the relative amplitude of the modulation does not decay (or grow). Therefore, we conclude that in principle 
the imprints of the cosmological dilution of the chiral plasma can be read from the gravitational wave spectrum, as expected from the discussion in Sec.~\ref{GW_sol}. If the primordial gravitational waves are produced in the early Universe and the Universe has experienced some epoch where non-zero chiral asymmetry of the fermions existed, we expect that such a gravitational birefringence is generated and remains until today. 
However, this in turn means that the memory of the intrinsic evolution of the chemical potential imprinted on the gravitational waves, which we discussed in Sec.~\ref{GW_f}, is considerably affected by the cosmological expansion.
Therefore, we need to sum up such contributions in order to give a precise prediction on the gravitational birefringence.
Although the summation would result in small correction and it would not be possible to detect such an effect within the current technology, the memory effect of the chiral imbalance may enable us to capture the dynamics of primordial thermal plasma through the future gravitational wave observation. 

\section{Discussion}
In this paper, we investigate the chiral gravitational effect in the time-dependent asymmetric fermion background by solving the equation of motion for the gravitational waves with the induced gravitational Chern-Simons term and evaluating its backreaction through the anomaly equation. As we discussed in Sec.~\ref{CS}, the formally covariant expression of the effective action~\eqref{gCS_f} enables us to consider the local anomalous contribution in such a situation, for high momentum gravitational waves $k \gtrsim \mu_5, T$. Compared to the CME, there are two specific features in the case of the chiral gravitational effect. First, as explicitly shown for the Minkowski spacetime in Sec.~\ref{CS}, the induced energy momentum tensor shows $\dot{\mu_5}$ dependence, which is absent in the case of CME. By solving the corrected equation of motion for the gravitational waves in Sec.~\ref{GW_f}, we have found that the gravitational birefringence takes place according to the evolution of $\dot{\mu_5}$ during the propagation. This implies that the dynamics of the  primordial chiral plasma is imprinted on the gravitational waves emitted before the generation of the chiral asymmetry through the chiral gravitational effect. We also consider the backreaction in Sec.~\ref{br_f}, where we have assumed that the time variation of $\mu_5$ is sourced by the external chirality-changing processes. We confirm that the memory, or the helicity accumulated in the gravitational waves, is dominated by the source contribution in the early times. We also find that the backreaction typically results in the oscillatory conversion between the chirality carried by fermions and the helicity of the gravitational waves. This oscillatory behavior, driven by the external chirality-changing process, is in contrast with the CME where the CPI is triggered by the constant $\mu_5$.   
The second feature is that cosmic expansion gives a non-trivial contribution due to the non-conformal nature of the graviton. 
As we demonstrate in Sec.~\ref{GW_FRW} and~\ref{br_FRW}, a diluting chemical potential also causes the gravitational birefringence and
drives oscillatory conversion between the chirality and the helicity as a backreaction. 
While the correction of the gravitational waves has the same scaling law as the free gravitational waves, $\propto 1/\eta$, the backreaction to the chiral plasma turns out to decay rapidly as $\propto 1/\eta^2$ and is highly suppressed by a small overall factor as in the Minkowski spacetime case. We would like to stress that there are no violent amplification of the gravitational field like CPI as long as the local description of the chiral gravitational effect is concerned, which ensures the backreaction to be negligible in our treatment. This difference originates from the different structure of the Chern-Simons term, or the symmetry of the system. While the Chern-Simons term causes the tachyonic instability for the spin-1 gauge field within the region of validity of the effective field theory, gravitational Chern-Simons term leads to the ghost instability for very high momentum modes. Since the effective field theory description breaks down in such a regime, we restrict our study in the scale much below the Planck or the cut-off scale where the Chern-Simons term correction can be perturbatively included, and reasonably avoid the appearance of %the 
ghost modes.
 
Let us make a qualitative discussion on the chiral gravitational effect in cosmology. If one considers some scenarios of baryogenesis like the GUT baryogenesis, for example, $\mu_5$ can be generated at the time of the baryogenesis and eventually decays when the Yukawa interaction gets in equilibrium~\cite{Campbell:1992jd,Kamada:2018tcs,Domcke:2020quw}. In this case, non-trivial time dependence on $\mu_5$ is given by the generation, decay, and the cosmological dilution. The memory of the chiral imbalance imprinted on the gravitational waves is determined by the combination of those contributions. As we demonstrated in Sec.~\ref{br_f} and Sec.~\ref{br_FRW}, the backreaction can be neglected 
for the propagation of the gravitational waves. 
It would enable us to investigate the memory effect of the  gravitational waves for various models of the baryogenesis or leptogenesis without complicated discussions. Such imprints of the primordial chirality violation in the plasma might be observed by the recently proposed high frequency gravitational wave detectors~\cite{Cruise:2012zz, Ito:2019wcb, Domcke:2020yzq, Aggarwal:2020olq} since the Chern-Simons term-like local contribution we investigated here becomes dominant for the high momentum region $k \gtrsim T$.

We would like to add a comment on the possibility that this memory effect in gravitational wave might become the probe of the dark sector. Since gravity couples to the any kinds of matter fields, the chiral gravitational effect can take place due to the chiral flipping process even in the dark sector. Although such effects would be too small to be detected with present gravitational detectors, it is worth investigating such chiral gravitational waves in detail. This possibility motivates us to make a further study of the chiral gravitational effects including non-local effects in general backgrounds, which we do not address in this work.

Finally, we would like to make a comment on the limitation of our study and the possible extension of it. 
First, we considered monotonic plane gravitational waves as the incoming wave. 
In more realistic case, especially for the application to cosmology, 
we need to take into account the stochasticity of the gravitational waves 
and the effect of their spectrum. 
This would become important when we also take into account the spatial dependence 
of the chiral chemical potential, which is neglected in this study. Moreover,
as we mentioned in Sec.~\ref{CS}, induced gravitational Chern-Simons term cannot reproduce the non-local contributions to the anomalous dispersion, which includes, for example, the thermal correction derived in the study of static thermal background~\cite{Manes:2012hf,Jensen:2012kj}. Although the Chern-Simons term and the spurious field enable us to appropriately take into account the time dependence of chemical potential of the background, the anomalous dispersion of the gravitational waves arising from this term becomes leading only when the conditions $\mu_5 \lesssim T \lesssim k$ are satisfied. If one considers the low momentum region $k \lesssim T$ instead, where the ground-based and space-born gravitational detector become sensitive, thermal corrections would dominate the anomalous dispersion relation of the gravitational waves. In this sense, extensions of our study to include thermal non-local corrections properly is necessary to probe the chiral thermal plasma in the early Universe by the present and planned gravitational wave detectors. However, incorporating thermal correction into our formally covariant effective Lagrangian is not so straightforward since the thermal effect associated to the global nature of background appears as the non-local contribution, as we mentioned in Sec.~\ref{CS}. To achieve such an extension, we need to take different approaches beyond the covariant local Lagrangian formalism we have taken. We leave such extensions for future study.

\section*{Acknowledgments}
The authors thank Naoki Yamamoto for useful discussions. This work was partially supported by JSPS KAKENHI Grant-in-Aid for Scientific Research (C) JP19K03842 (KK) and Innovative Area 19H04610(KK). JK is supported by JSPS KAKENHI, Grant-in-Aid for JSPS Fellows 20J21866 and research program of the Leading Graduate Course for Frontiers of Mathematical Sciences and Physics (FMSP). YY is supported by JSPS KAKENHI, Grant-in-Aid for JSPS Fellows JP19J00494.

\bibliographystyle{JHEP}
\bibliography{draft.bib}
\end{document}